\documentclass[twocolumn,english,showpacs,pra]{revtex4}
\usepackage[T1]{fontenc}
\usepackage[latin9]{inputenc}
\usepackage{amsmath}
\usepackage{amssymb}
\usepackage{graphicx}
\usepackage{esint}

\makeatletter

\providecommand{\tabularnewline}{\\}

\@ifundefined{textcolor}{}
{%
 \definecolor{BLACK}{gray}{0}
 \definecolor{WHITE}{gray}{1}
 \definecolor{RED}{rgb}{1,0,0}
 \definecolor{GREEN}{rgb}{0,1,0}
 \definecolor{BLUE}{rgb}{0,0,1}
 \definecolor{CYAN}{cmyk}{1,0,0,0}
 \definecolor{MAGENTA}{cmyk}{0,1,0,0}
 \definecolor{YELLOW}{cmyk}{0,0,1,0}
}

\usepackage{amsfonts}

\makeatother

\usepackage{babel}
\begin{document}

\title{Gain properties of dye-doped polymer thin films}

\author{I. Gozhyk$^{1}$, M. Boudreau$^{1,5}$, H. Rabbani Haghighi$^{2}$, N. Djellali$^{1}$,
S. Forget$^{2}$, S. Chenais$^{2}$, C. Ulysse$^{3}$, A. Brosseau$^{4}$,
R. Pansu$^{4}$, J.-F. Audibert$^{4}$, S. Gauvin$^{5}$, J. Zyss$^{1}$,
and M. Lebental,$^{1}$}

\affiliation{$^{1}$ Laboratoire de Photonique Quantique et Moléculaire, CNRS
UMR 8537, Institut d'Alembert FR 3242, Ecole Normale Supérieure de
Cachan, 61 avenue du président Wilson, F-94235 Cachan, France.\\
 $^{2}$ Laboratoire de Physique des Lasers, Universite PARIS 13 et
CNRS UMR 7538, 99 Avenue Jean-Baptiste Clement, F-93430 Villetaneuse,
France.\\
 $^{3}$ Laboratoire de Photonique et Nanostructures, CNRS UPR20,
Route de Nozay, F-91460 Marcoussis, France. \\
 $^{4}$ Laboratoire de Photophysique et Photochimie Supramoléculaires
et Macromoléculaires, CNRS UMR 8531, Institut d'Alembert FR 3242,
Ecole Normale Supérieure de Cachan, F-94235 Cachan, France.\\
 $^{5}$ Groupe de recherche sur les couches minces et la photonique,
Département de physique et d'astronomie, Université de Moncton, Moncton,
NB, Canada E1A 3E9.}
\begin{abstract}
Hybrid pumping appears as a promising compromise in order to reach the much coveted goal of an electrically pumped organic laser. In such configuration the organic material is optically pumped
by an electrically pumped inorganic device on chip. This engineering
solution requires therefore an optimization of the organic gain medium
under optical pumping. Here, we report a detailed study of the gain
features of dye-doped polymer thin films. In particular we introduce
the gain efficiency $K$, in order to facilitate comparison between
different materials and experimental conditions. The gain efficiency
was measured with various setups (pump-probe amplification, variable
stripe length method, laser thresholds) in order to study several
factors which modify the actual gain of a layer, namely the confinement
factor, the pump polarization, the molecular anisotropy, and the re-absorption.
For instance, for a 600 nm thick 5 wt\% DCM doped PMMA layer, the
different experimental approaches give a consistent value $K\simeq$
80 cm.MW$^{-1}$. On the contrary, the usual model predicting the
gain from the characteristics of the material leads to an overestimation
by two orders of magnitude, which raises a serious problem in the
design of actual devices. In this context, we demonstrate
the feasibility to infer the gain efficiency from the laser threshold
of well-calibrated devices. Besides, temporal measurements at the
picosecond scale were carried out to support the analysis.
\end{abstract}


\maketitle

\section{Introduction}

Photonic technology based on organic materials has continuously progressed
over the last decades \cite{livre-seb}. Organic diodes (so-called
OLEDs) have developed into an industrially viable domain, whereas
polymer based integrated optical devices \cite{myata,nalwa,shinar}
have matured into robust alternatives to semiconductor devices. Among
their advantages are the possibility for flexible substrates \cite{koike},
the quasi-unlimited versatility of materials \cite{materiaux}, and
a more favorable bio- or chemical compatibility \cite{chao}.\\
 However, for more than twenty years, the direct electrical excitation
of stimulated emission in organic semiconductors remains a major challenge
\cite{baldo,Ifor_natphot}, although solid-state organic lasers demonstrated
their high potential under optical pumping \cite{ifor review,sebastiens pol. international}.
In fact, electrically-pumped organic materials exhibit several extra
losses mechanisms (triplet-induced problems \cite{baldo el pump,Forrest STA},
charge-induced absorption \cite{tessler APL99,kozlov cur abs} and
absorption at metal contacts), which grow rapidly with the current
\cite{Ifor_natphot}, creating a negative feedback loop \cite{ifor review,note:limite-seuil}.\\

\begin{figure}[htb]
\centerline{\includegraphics[width=8.3cm]{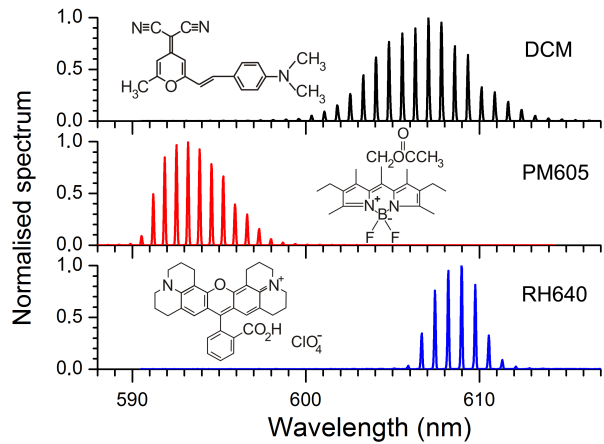}}
\caption{Molecular structures of the laser dyes studied in this article, and
normalized emission spectra of a 150 $\mu$m Fabry-Perot like micro-laser
doped with the corresponding dye.}
\label{fig:spectra}
\end{figure}

Therefore the excitation of organic materials via an inorganic electroluminescent
pump \cite{baldo el pump} is now considered as more realistic and
thus as a highly promising approach, since it allows to achieve an
\emph{indirect} electrical pumping of the gain medium. Lasing in such
hybrid system was successfully demonstrated under various pumping
sources: micro-chip lasers \cite{voos micorchip}, inorganic laser
diodes \cite{Riedl dfb InGaN,samuel brag diodepump}, and even incoherent
LEDs \cite{samuel hybrid LED}. Indirect electrical excitation requires
to carefully examine optical pumping to take into account the specificities
of these new pump sources and of organic materials. Plasmon-assisted
organic emitters and especially spasers (plasmonic lasers) would also
take benefit of such a study, since their gain medium often involves
organic materials \cite{Noginov spaser,gather-plasmon,shalaev-spaser}.\\
 It is relatively easy to achieve lasing with laser dyes, since these
small molecules are commercially available and provide stimulated
emission in a great variety of host matrices. However, quantifying
their lasing features is much more difficult, in particular predicting
the linear gain and lasing thresholds. In this paper, we present a general perspective on the gain properties of dye-doped thin films,
using several complementary approaches, such as threshold
measurements, linear gain measurements, temporal measurements, and simulations.
In order to provide guidelines for the comparison of organic materials,
a systematic description of gain properties is introduced. In fact this field has remained largely unexplored, apart
from a few works on particular concentrations of specific dye molecules
\cite{lam,alq3-dcm laser,Lu,costella pm567}. For this purpose, our
experimental test-beds are dye-doped polymer thin-film lasers based
on commercial laser dyes, namely DCM, Rhodamine 640 (RH640) and Pyrromethene
605 (PM605), embedded in a passive matrix, in conventional configurations such as Amplified Spontaneous
Emission (ASE) and Fabry-Perot like cavities, see Fig.\ref{fig:spectra}.
The method proposed herein aims at facilitating the engineering of photonic
devices and providing a tool for material preselection.\\

The article is organized as follows. First, Section \ref{sec:configuration-experimentale}
presents the samples and the optical characterization setups. Then section \ref{sec:gain} provides a general
description of the gain, and introduces the relevant parameters, which
are then studied in details in the following Sections. The specificities
of the geometry are taken into account via the confinement factor
and the modal gain, which are discussed and measured in Sec.\ref{sec:modal-gain}.
Based on these measured amplification properties, Sec.\ref{sec:seuil}
provides a comparison between the experimental laser thresholds and
the estimated values in the testbed configuration of Fabry-Perot microlasers.
The good agreement and the reliability of this method stress the advantage
to calibrate gain measurements on laser threshold of well-controlled
devices. Then, in Sec.\ref{sec:pola} the impact of the pump polarization
on gain is accounted for through a model based on fluorescence anisotropy
in good agreement with experiments. Spectral features influencing
the evaluation of gain are finally detailed in Section \ref{sec:Mazumder}.
The last section, Sec.\ref{sec:pompe-sonde}, deals with the direct
measurement of the bulk gain by amplification of a probe beam, which
leads to an experimental value consistent with the previous ones,
contrary to theoretical estimations, which overestimate the gain by
two orders of magnitude.\\
 Several appendixes provide additional descriptions of this system.
App.\ref{sec:lifetime} and \ref{sec:aggregats} deal with time-resolved
experiments, while App.\ref{sec:rho} and \ref{sec:section-efficace}
set some definitions on fluorescence anisotropy and cross sections
of laser dyes.

\section{Experimental configurations}

\label{sec:configuration-experimentale}

\begin{figure}[htb]
\centerline{\includegraphics[width=8.3cm]{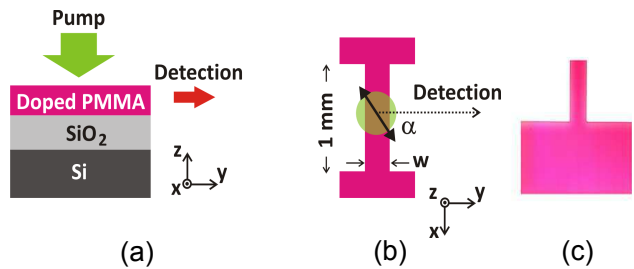}}
\caption{Usual sample configuration, as described in Sec.\ref{sec:experience-usual}: (a) The sample stack,
not to scale; (b) In-plane scheme of a ribbon (ie. Fabry-Perot like) resonator. The pumped
part is indicated by a green disk. The angle between the polarization
of the pump beam (double arrow) and the direction of the detection
($y$ axis) is called $\alpha$. (c) Optical microscope photo in real
colors of a $w=165$ $\mu$m PMMA-PM605 ribbon micro-laser (partial
view - bottom part).}
\label{fig:setup}
\end{figure}

The study reported in this article is focused on actual optically-pumped
organic lasers made of a single polymer thin film, as presented in
Fig.\ref{fig:setup}. In order to get a survey on their gain properties,
several complementary experiments were carried out on different sample
types, adapted to each experimental configuration. This Section gathers
information on the various experimental setups used through the article
and their typical samples. First the lasing characterization setup,
the fabrication of samples and their geometry are described. Then
we detail the specificities of the samples used for amplification
measurements in Sec.\ref{sec:pompe-sonde}. Finally, both setups used
for time-resolved experiments are briefly discussed, with more details
given in App.\ref{sec:lifetime}.

\subsection{Laser experiments}

\label{sec:experience-usual}

The typical layer stack is presented in Fig.\ref{fig:setup}a. It
is made of a 600 nm thick PMMA layer (poly(methyl metacrylate), purchased from
MicroChem) doped with a laser dye (5 wt\%), and spin-coated on
a commercial 2 $\mu$m Si0$_{2}$/Si wafer. For all experiments except
amplification measurements (Sec.\ref{sec:pompe-sonde}), we used
PMMA with a molecular weight of 495,000 at a concentration of 6
wt\% in anisole. The laser dyes were bought at Exciton and used as
received.\\

The polymer layer can be patterned by electron-beam lithography, which
generates arbitrary cavity shapes with a
nanoscale etching quality \cite{lebental-matsko}. In Sec.\ref{sec:seuil}
and \ref{sec:pola}, we focus on the simplest case of Fabry-Perot
resonators, in order to compare threshold measurements with theoretical
predictions.\\
 The sample is pumped from the top (along the $z$ axis, see notations
in Fig.\ref{fig:setup}) by a frequency doubled Nd:YAG laser (532
nm, 500 ps, 10 Hz). Then the emitted intensity is collected at the
sample edge, in the layer plane ($xy$ plane). In fact, the index
contrast between the silica and the polymer layer allows for the confinement
of the electromagnetic fields within the layer plane ($xy$ plane), while the thickness of the film is
chosen to be of the order of the emission wavelength to get a single excitation along $z$ \cite{lebental-spectre}.\\
 The energy and the linear polarization of the pump beam are controlled
independently with a standard combination of linear polarizers and
half-wave plates. The polarization of the pump beam lies within the layer
plane ($xy$ plane) and its orientation in this plane is defined by the
angle $\alpha$ with respect to the $y$-axis (see Fig.\ref{fig:setup}b),
namely $\alpha=0^{\circ}$ if the polarization is oriented along the
$y$-axis, and $\alpha=90^{\circ}$ if the pump polarization is along
the $x$-axis. The influence of $\alpha$ is discussed in Section \ref{sec:pola}.
Unless otherwise specified, a pump polarization with $\alpha=90^{\circ}$ is being considered.

\subsection{Amplification experiments}

\label{sec:experience-amplification}

In Sec.\ref{sec:pompe-sonde}, we report the measurements of the linear
material gain $g_{mat}$ by means of a pump-probe experiment. The
pump laser was identical to the above-described experiment. The probe
beam at 594 nm was sent to measure the amplification by the excited
dye-doped polymer layer. Probe wavelength was chosen in the spectral
region with high dye emission cross-section and negligible dye absorption. The setup is further detailed in Sec.\ref{sec:pompe-sonde}.

In order to get a longer interaction path between the probe beam and
the excited film, the thickness and dye concentration of the dye-doped
polymer layer as well as its substrate were changed. The sample used
in amplification experiment was a 18 $\mu$m PMMA layer (molecular
weight 950,000, 15 wt\% in anisole) doped with 1.4 wt\% DCM, and spin-coated
on a glass substrate.

\subsection{Time-resolved experiments}

Since organic emitters can return from the excited energy state to
the ground state through different pathways \cite{sebastiens pol. international,lakowicz,valeur},
the experimental analysis of any emission from organic semiconductors
is not complete without the characterization of the emission dynamics.
For this purpose, we used two different experimental configurations.
\\
 First, a streak camera was recorded the temporal behavior of
usual samples (see Subsec.\ref{sec:experience-usual}), and shows
the difference between fluorescence, ASE, and laser dynamics. The
setup is presented in App.\ref{sec:lifetime}. The main results are
gathered in Fig.\ref{fig:ase-fluo-lifetime}.\\
 Then the decay of fluorescence was measured with a Fluorescence-Lifetime
Imaging Microscope (FLIM) described in App.\ref{sec:aggregats}. The
samples were made of a 600 nm thick PMMA layer spin-coated on a glass
substrate and doped with various concentrations of laser dyes, in
order to evidence the influence of quenching, as discussed in Sec.\ref{sec:pompe-sonde}.

\section{Gain in thin films}

\label{sec:gain}

Gain properties of active media are of utmost importance for the design
of photonic devices. However, the usual gain terminology is not appropriate
for comparing amplification in different organic materials under different
pumping geometries. In this section, we analyze the commonly used
definitions of gain and their limitations, and introduce an alternative and more relevant way to account for gain in organic thin films. \\
 Two parameters are used in the literature to describe amplification:
\emph{material} and \emph{modal} gain. Both describe an average
growth rate of electromagnetic flux per unit medium length (in cm$^{-1}$, \cite{note:gain-decibel}), but in different systems. The \emph{material} gain represents
the gain in bulk, whereas the \emph{modal} gain describes amplification
in the exact thin-film geometry, accounting for transverse overlap
of the material gain ($g_{mat}$) with the pump and propagating mode
profiles ($\Gamma$) inside the layer \cite{dal Negro}:

\begin{equation}
g_{mod} =\Gamma\,g_{mat}-\alpha_{mod}\label{eq:gmodal-material}
\end{equation}
Modal losses $\alpha_{mod}$ should not be confused with material
(bulk) losses $\alpha_{mat}$, that will be described below. Modal losses encompass all the losses that emerge due to diffraction or scattering at the layer interfaces, i.e. losses that are not present in a homogeneous bulk medium.  \\
 In the literature, experimental gains are generally obtained in thin-film
configuration with the ``Variable Stripe Length'' technique (VSL)
\cite{dal Negro}. But, as a matter of fact, these are \emph{modal}
gain values and are therefore strongly influenced by the sample geometry
(layer thicknesses and refractive indexes), as well as experimental
conditions (eg. wavelength and orientation of the pumping beam). Thereby,
despite a great number of publications on gain in organic materials,
a systematic comparison of gain properties is still lacking due to
variations of experimental configurations. It is possible to account
for the impact of the system geometry through the estimation of the
confinement factor $\Gamma$. However, this demands supplementary
information, in particular the absorption and stimulated emission
spectra of the material under study. An example is provided in Section
\ref{sec:confinement}.\\

To reach the laser threshold, the challenge lies in getting
the material gain $g_{mat}$ greater than the losses.
\emph{A priori}, the gain of the bulk material $g_{mat}$ is not a
constant, it depends on several parameters including the wavelength
and pump intensity $I_{p}$. Gain in inorganic semiconductors is known
to vary logarithmically with the carriers density, which in fact is
proportional to the pump intensity, thus leading to: $g_{mat}\propto\ln\left(I_{p}\right)$
\cite{coldren}. However, early works on dye lasers state that amplification
in such medium is proportional to the density of excited molecules
\cite{peterson}, which in its turn depends linearly on the density
of absorbed photons. This leads to the linear dependance of the gain
on pump intensity in an intensity range limited by saturation and
non-linear effects:
\begin{equation}
g_{mat}  =KI_{p}-\alpha_{mat}\label{eq:gKI}
\end{equation}
where $\alpha_{mat}$ gathers various losses mechanisms, some independent
on pump intensity (traps, for instance), and some varying with the
pump intensity, like polarons or excited states.\\
 The linear coefficient $K$, referred hereafter to as \textit{gain efficiency,} provides an alternative gain description. We found a confirmation
of expression (\ref{eq:gKI}) in some publications \cite{Lu,costella pm567,Tagaya 2,McGehee},
listing modal gain at several pump intensity values. The inferred
gain efficiency $K$ is about 10\textsuperscript{1} to 10\textsuperscript{2}
cm.MW\textsuperscript{-1} for dye-doped systems \cite{Lu,costella pm567}
and about 10\textsuperscript{3} cm.MW\textsuperscript{-1} for conjugated
polymers \cite{McGehee}. The unit cm.MW$^{-1}$ reveals that we consider
the pump intensity and not the pump fluence, following the discussion
in Sec.2.6.4 of \cite{livre-seb}.\\
 A more comprehensive description of the material gain $g_{mat}$
includes spectral influence, since the absorption and emission cross-sections,
$\sigma_{a}$ and $\sigma_{e}$, depend significantly on the frequency.
The consequences on gain and lasing features are described in Section
\ref{sec:Mazumder}. A pragmatic evaluation of both cross sections
is presented in Appendix \ref{sec:section-efficace}.\\

In this section, we introduced the gain efficiency $K$ to describe
the amplification properties of the bulk material. It can be measured
by pump-probe experiment as illustrated in Sec.\ref{sec:pompe-sonde}.
However, the actual gain of a layer is modified by additional parameters:
the confinement factor $\Gamma$ (Sec.\ref{sec:modal-gain}), the
polarization properties (Sec.\ref{sec:pola}), and the spectral features
(Sec.\ref{sec:Mazumder}). As long as the saturation of absorption remains
negligible, expression (\ref{eq:gKI}) is a good approximation of
the material gain. In the next section, this linear dependance is
experimentally confirmed in a reasonable range of pump intensities.

\section{Modal gain $g_{mod}$}

\label{sec:modal-gain}

The Variable Stripe Length (VSL) technique is commonly used as a basic
method to measure the gain in a thin film device \cite{dal Negro,vsl-sastre}.
In fact, it gives access to the modal gain $g_{mod}$, which depends
on the overlap between the pumped region and the laser field. In this
Section, we first measure the modal gain of our usual layer stack
described in Sec.\ref{sec:experience-usual}. Then a general formulation
of the confinement factor $\Gamma$ is proposed and calculated for
this specific case, in order to evidence the influence of the sample
geometry and experimental configuration.

\subsection{VSL method}

\label{sec:vsl}

\begin{figure}[htb]
\centerline{\includegraphics[width=7.5cm]{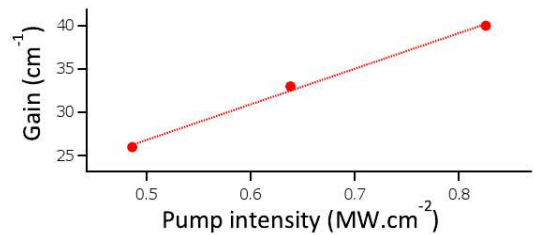}}
\caption{Modal gain versus pump intensity for a 600 nm thick 5 wt\% DCM-doped
PMMA layer on a 2 $\mu$m SiO$_{2}$/Si layer, measured by VSL method.}
\label{fig:vsl-hadi}
\end{figure}

The modal gain $g_{mod}$ of the usual layer stack described in Sect.\ref{sec:experience-usual}
was measured by means of the Variable Stripe Length (VSL) technique.
The pump beam was shaped as a rectangle of fixed width (about 300
$\mu$m) and variable length $L$. Then the emitted intensity was collected
at the sample edge, in the layer plane ($xy$ plane), and the modal
gain can be inferred from its variation versus the length $L$ at
a fixed pump intensity. Further details on this experiment can be
found in \cite{hadi}. As shown in Fig.\ref{fig:vsl-hadi} for a DCM
doped PMMA layer, the modal gain varies linearly with the pump intensity,
which evidences that Eq. (\ref{eq:gKI}) is sensible. The gain efficiency
was then inferred from a linear fit: $\Gamma K=41\pm2$ cm.MW$^{-1}$. This
value depends on the specificities of the layer stack. As shown in
the next subsection, the confinement factor $\Gamma$ can be numerically
calculated to estimate the influence of geometrical parameters.\\
 Even if this value of $K$ is consistent with other experimental
measurements (see below, in particular Sec.\ref{sec:seuil} and Sec.\ref{sec:pompe-sonde}),
the VSL technique is known to be prone to artifacts \cite{dal Negro,vsl-artefact}.

\subsection{Confinement factor}

\label{sec:confinement}

The confinement factor $\Gamma$ was introduced in Eq. (\ref{eq:gmodal-material})
in order to account for the overlap of the material gain $g_{mat}$ with
the pump $E_{p}$ and lasing $E_{l}$ mode profiles \cite{elena-jqe}.
Therefore it depends on the specific lasing mode $E_{l}$ which is
considered. Eq. (\ref{eq:gmodal-material}) is an abridged version
of an expression, which involves the integration of the spatial profiles
of the fields and the gain efficiency over the volume of the gain
layer. For the sake of simplicity, we assume that the field profiles
are uniform within the sample plane (along $x$ and $y$-axes, see
Fig.\ref{fig:setup}), and consider only field variations along the
layer thickness ($z$-axis), then integration occurs only along the
thickness of the gain layer $h$:

\begin{equation}
\Gamma=\frac{\int_{h}dz\,\tilde{f}(z)|E_{l}(z)|^{2}}{\int_{\mathbb{R}}dz\,|E_{l}(z)|^{2}}\label{eq:confinement-general}
\end{equation}

where $\tilde{f}(z)$ is the normalized function describing the distribution
of the excited molecules along the $z$ axis.

In some experiments, the gain is uniform within the layer, which means
$\tilde{f}(z)$=1 inside (for instance, no concentration variation
and excitation homogeneously distributed within the layer) and null
outside the layer. In this specific case, the confinement factor is
defined by the ratio of the laser field overlapping the gain layer
\cite{corzine} :
\begin{equation}
\Gamma=\frac{\int_{h}dz\,|E_{l}(z)|^{2}}{\int_{\mathbb{R}}dz\,|E_{l}(z)|^{2}}\label{eq:confinement-corzine}
\end{equation}
The confinement factor varies then between 0, when there is no overlap
between the laser field and the gain region, and 1, when the whole
field is located within the excited layer.\\

\begin{figure}[htb]
\centerline{\includegraphics[width=8.3cm]{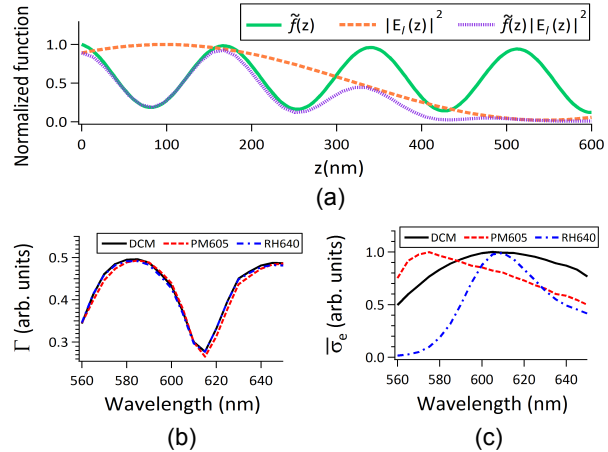}}
\caption{(a) Distribution of $\tilde{f}(z)$ for a 5 wt\% DCM doped PMMA layer pumped
at 532 nm, from Eq. (\ref{eq:f-tilde}), distribution of normalized $|E_{l}(z)|^{2}$
at 600 nm, and plot of $\tilde{f}(z)|E_{l}(z)|^{2}$. $z=0$ stands for the air-polymer interface. The pump is only slightly absorbed by the DCM doped layer. (b) Confinement factors $\Gamma$ calculated from Eq. (\ref{eq:confinement-general}) in a 600 nm thick 5 wt\% dye-doped PMMA layer for DCM, RH640 and PM605 dyes. (c) Normalized emission
cross-sections, inferred from Eq.(\ref{eq:section-emission}).}
\label{fig:facteur confinement}
\end{figure}

However, in most configurations, the pump profile or the distribution
of the material gain is not uniform within the gain layer, and this
distribution must be taken into account. In this case, we consider
the normalized function $\tilde{f}$ defined by:
\begin{equation}
\tilde{f}(z)=\frac{\eta(z)|E_{p}(z)|^{2}}{Max[\eta|E_{p}|^{2}]}\label{eq:f-tilde}
\end{equation}
where $\eta(z)$ is the density of emitters over the layer thickness ($z$ axis) and $Max[\eta|E_{p}|^{2}]$
is the maximum value of the function $\eta(z)|E_{p}(z)|^{2}$. As
$\tilde{f}$ varies between 0 and 1, the emission field acts as
an envelope for $\tilde{f}(z)|E_{l}(z)|^{2}$ in Eq. (\ref{eq:confinement-general}). As a result, the upper
limit of $\Gamma$ becomes inferior to 1 and depends on the overlap
between the emission profile and $\tilde{f}$. Moreover, Eq. (\ref{eq:confinement-general})
leads to Eq. (\ref{eq:confinement-corzine}) for a uniform gain,
as expected.\\
 The confinement factor can be calculated for a given geometry, for
instance for the configuration of the slab waveguide presented in Fig.\ref{fig:setup}a.
For the pump field, we consider a standing wave problem in a multi-layered
system: the dye doped polymer (refractive index $n$ =1.54 and thickness
$h=600$ nm) and the SiO$_{2}$ layer ($n$ =1.46, $h=2\,\mu$m) placed
between the semi-infinite medium (air, $n$=1) and the substrate (Si,
$n$ =4.14). For the propagating laser field, we use the model of
the effective index described in \cite{lebental-spectre}, and consider
the first vertical excitation with TE polarization, which means that
the electric field lies in the plane of the layer. Actually, for ASE
and Fabry-Perot microlasers, the laser emission is mostly TE polarized
\cite{PRA Iryna}.\\
 The confinement factor was numerically estimated based on Eq. (\ref{eq:confinement-general}).
The pump profile $|E_{p}(z)|^{2}$ varies within the layer due to
absorption. Here, we consider a fixed pump wavelength, $\lambda_{p}=532$
nm, in accordance with the experiments described in this article. Then, the
confinement factor $\Gamma$ depends on the wavelength because of
two different physical terms, namely the profile of the laser field
$|E_{l}(z)|^{2}$, and the emission cross-section via $\tilde{f}$.
Fig.\ref{fig:facteur confinement}b evidences that $\Gamma$ remains
close to 0.5 for the specific geometry considered here. In this configuration, the
variation of the confinement factors with the wavelength is quite similar for the three dyes (see Fig.\ref{fig:facteur confinement}b), which means that the emission wavelength in such devices will be defined
by the spectral shape of the emission cross-section (Fig.\ref{fig:facteur confinement}c).

\begin{figure}[htb]
\centerline{\includegraphics[width=8.3cm]{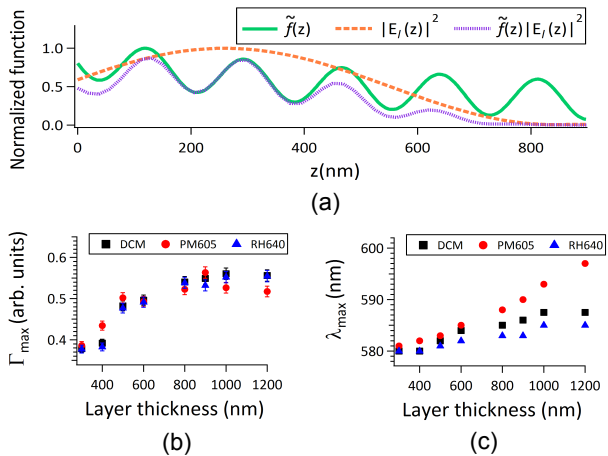}}
\caption{(a) Distribution of $\tilde{f}(z)$ for a 5 wt\% PM605 doped PMMA layer pumped
at 532 nm, from Eq. (\ref{eq:f-tilde}), distribution of normalized $|E_{l}(z)|^{2}$
at 590 nm, and plot of $\tilde{f}(z)|E_{l}(z)|^{2}$. $z=0$ stands for the air-polymer interface.
(b) Maximum of the confinement factor for the three dyes versus the thickness of the gain layer. (c) Wavelength corresponding to the maximum value of the confinement factor versus the thickness of the layer.}
\label{fig:facteur confinement bilan}
\end{figure}

Finally, Fig.\ref{fig:facteur confinement bilan} presents the influence
of the thickness of the dye-doped PMMA layer on the confinement factor
$\Gamma$. In general, the increase of the layer thickness up to some
limit value (around 900 nm) results in an increase of $\Gamma$, and
thus of the gain. Actually the absorption of the pump leads to a
non-homogeneous excitation profile within the layer (see Fig.\ref{fig:facteur confinement}a).
Increasing the layer thickness results in the decay of the pumping
field magnitude towards the substrate (Fig.\ref{fig:facteur confinement bilan}a).
In case of moderate decay of the pumping field magnitude within the layer,
the overlap of the excitation and emission patterns improves. However,
when the gain layer becomes strongly,
the excitation profile decays faster than the emission profile thus decreasing
the overlap. The emission wavelength is also impacted by the layer
thickness, as shown in Fig.\ref{fig:facteur confinement bilan}c.

\subsection{Conclusion on modal gain}

In conclusion, the amplification properties are strongly influenced
by the exact geometry of the system and the confinement factor is
an appropriate tool to take into account such parameters. But, to
have access to the gain of the actual device, the input of the material
gain $g_{mat}$ is necessary. It can be inferred from VSL experiments,
if $\Gamma$ is known. For instance, in the layer stack considered here, $\Gamma$ is about 0.5, which means that the gain efficiency inferred from VSL experiments was in fact: $ K=82\pm4$ cm.MW$^{-1}$. A good agreement is evidenced in Sec.\ref{sec:pompe-sonde}.
However VSL experiments are prone to artifacts \cite{dal Negro,vsl-artefact},
and it is sometimes more convenient and reliable to determine the
gain from laser thresholds of well-controlled devices, as shown in
the next Section.

\section{Threshold estimate}

\label{sec:seuil}

Once the gain efficiency $K$ is known, it can be used to quantitatively
predict the lasing threshold intensity of a specific device. Reciprocally,
$K$ can be inferred from threshold measurements in well-controlled
configurations. This Section deals with both aspects of this issue,
using simple and well-known Fabry-Perot resonators.

The sample type and the characterization setup are described in Sec.\ref{sec:experience-usual}. The gain efficiency $K$ measured
by VSL in Sec.\ref{sec:vsl} should then be still valid. To obtain
a Fabry-Perot resonator, a ribbon shape was chosen. Ribbon cavities
were fabricated with different widths $w$ ranging from 100 $\mu$m
to 200 $\mu$m. As shown in Fig.\ref{fig:setup}bc, there is no border
in the $x$-direction, in order to prevent back reflection along the
$x$-axis, and therefore to avoid mode competition, which would modify
the lasing thresholds. We then expect Fabry-Perot modes propagating
along the $y$-axis.

\begin{figure}[htb]
\centerline{\includegraphics[width=8.3cm]{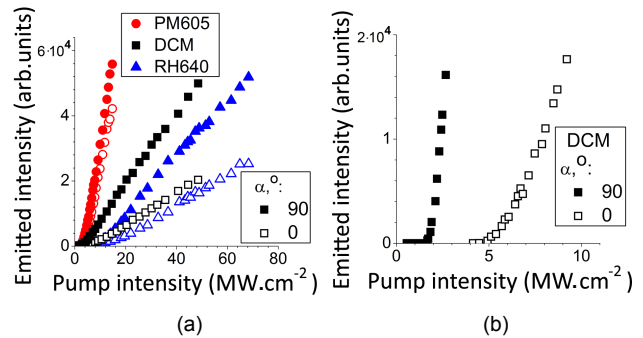}}
\caption{Influence of the pump polarization $\alpha$ on the emission. Emission
intensity versus pump intensity in the case of (a) ASE, (b) a 165 $\mu$m DCM Fabry-Perot micro-laser.} \label{fig:ase-FP}
\end{figure}

\begin{figure}[htb]
\centerline{\includegraphics[width=8.3cm]{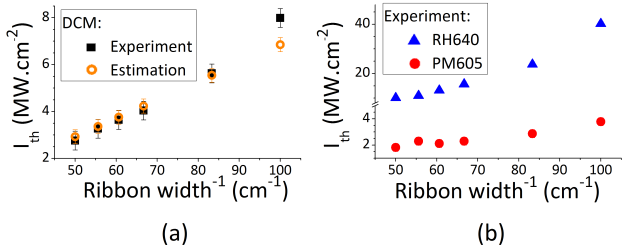}}
\caption{Experimental threshold intensities versus $1/w$ at $\alpha=90^{\circ}$
with different laser dyes: (a) DCM (together with estimated values
from Eq. (\ref{eq:seuil}), using $\Gamma K$=41 cm.MW$^{-1}$ inferred in
Sec.\ref{sec:vsl}), (b) PM605 and RH640.}
\label{fig:seuil_estimation}
\end{figure}

The ribbon resonator was pumped partially, as presented in Fig.\ref{fig:setup}b.
Here, we consider the most favorable case, where the polarization
of the pump beam is perpendicular to the $y$-axis, ie. $\alpha=90^{\circ}$,
like in VSL experiments described in Sec.\ref{sec:vsl}. Typical Fabry-Perot
spectra are presented in Fig.\ref{fig:spectra}. Their Free Spectral
Range (FSR) indicates that the lasing modes correspond indeed to Fabry-Perot
resonances \cite{lebental-spectre}. Fig.\ref{fig:ase-FP}b shows
typical I-I plots, where the threshold intensity $I_{th}$ is
identified as a change of slope with a precision of about 0.1 MW.cm$^{-2}$ \cite{note:seuil-mesure}. Then the experimental thresholds for the three dyes are gathered in Fig.\ref{fig:seuil_estimation}.

To predict the threshold intensity, we assume that the stationary
regime is reached. This assumption is validated \emph{a posteriori}
thanks to the good agreement with experiments. In the stationary regime,
the threshold is determined by the modal gain (and thus the pump intensity)
sufficient to compensate the losses:

\begin{equation}
re^{g_{th}2w} =1\label{eq:gain-pertes}
\end{equation}
with $r$ standing for the total losses in the system. For the sake
of simplicity, we only consider refraction losses (the major factor
of losses in our system): $r=R^{2}$, where $R=\left(\frac{n-1}{n+1}\right)^{2}\simeq0.04$
is the reflection coefficient at the boundary ($n_{cavity}=n\simeq1.5$
/ $n_{air}=1$). Pump intensity at the threshold is then expected
to be linear with $1/w$:
\begin{equation}
I_{th}=-\frac{lnR}{\Gamma K}\frac{1}{w}+\frac{\alpha_{mod}}{\Gamma K}\label{eq:seuil}
\end{equation}
using $g_{th}=g_{mod}=\Gamma K\,I_{th}$.
As reported in Fig.\ref{fig:seuil_estimation}, pump intensities at
threshold show indeed a linear behavior with $1/w$. Moreover, a quantitative
agreement with Eq.(\ref{eq:seuil}) is demonstrated for DCM in Fig.\ref{fig:seuil_estimation}a,
using $R=0.04$ and $\Gamma K$=41 cm.MW$^{-1}$, without adjusted parameter.\\
 This agreement has several consequences. (i) The assumption of
a stationary regime should be valid. (ii) Spatial hole-burning does
not influence the thresholds, whereas dye molecules lead to an homogeneous
gain and the spectra are multimode, even at threshold. (iii) Only
the losses due to refraction are taken into account in Eq. (\ref{eq:seuil}).
Hence the quantitative agreement means that the losses due to diffraction
at the cavity edges do not modify the thresholds, as evidenced in
\cite{lozenko}, and that the Fresnel coefficient for an infinite
wall $R$ does reproduce correctly the refraction at the boundary,
even if the cavity thickness scales with the wavelength.

Fig.\ref{fig:seuil_estimation} also shows departure from a linear
behavior for the smallest cavities (around 100 cm$^{-1}$).
More intricate active phenomena occur in these cases and will be discussed
in Sec.\ref{sec:Mazumder}.

Reciprocally, Eq.(\ref{eq:seuil}) can be used to infer the gain efficiency
for the three dyes. The fit results are given in Table \ref{tab:K}.
At first view, it would seem that PM605 is a more efficient laser dye than
DCM and RH640. However, the conclusion is intricate, since the pump
wavelength and the dye concentration were fixed, whereas
a proper comparison would require a specific optimization for each laser dye.

\begin{table}
{\small{{\centering \caption{{\small{{\label{tab:K} Comparison of gain efficiencies $\Gamma K$ for
the three dyes inferred by fitting the experimental data of Fig.\ref{fig:seuil_estimation}
with Eq. (\ref{eq:seuil}), using $R=0.04$.}}}}
}}{\small \par}
{\small{}}}{\small \par}
\centering %
\begin{tabular}{c|ccc}
\hline
Dye  & DCM  & RH640  & PM605\tabularnewline
\hline
$\Gamma K$ {[}cm MW$^{-1}${]}  & 42 $\pm$ 4  & 9 $\pm$ 1  & 81 $\pm$ 2\tabularnewline
\hline
\end{tabular}
\end{table}

To summarize, we evidenced that the gain efficiency $K$ inferred
by the VSL technique in Sec.\ref{sec:vsl} is consistent with threshold
measurements of testbed Fabry-Perot micro-lasers. Moreover, predictions
can be made with good precision (less than 10 $\%$ of uncertainty).
Instead of measuring the gain by VSL or pump-probe experiments, it
is hence easier and more reliable to infer $K$ from the threshold
intensity of well-controlled laser testbeds, such as Fabry-Perot microlasers, the confinement factor $\Gamma$ being calculated independently.

\section{Influence of the pump polarization}

\label{sec:pola}

Absorption and emission of light in organic semiconductors are known
to be strongly polarization-dependant \cite{valeur,lakowicz,PRA Iryna}.
For instance, pump polarization can modify ASE intensity and laser
threshold \emph{up to a factor of three}. It is thus essential to
take this issue into account. Therefore this section focuses on the
influence of the pump polarization on gain. A model is provided in
quantitative agreement for the three laser dyes.

Due to the molecular structure of organic materials, amplification
is in general \emph{anisotropic} in such media \cite{PRA Iryna}.
In an approximate model, dye molecules in a polymer matrix can be considered
as fixed and non-interacting dipoles. Each molecule absorbs preferentially
along the direction of its absorption dipole, and emits a fluorescent
photon according to its emission dipole. Both dipoles depend on the
geometry of the molecular structure. This explains, that an ensemble
of dye molecules can emit fluorescence in specific directions (monitored
by the pump polarization), even if they are isotropically distributed.
This phenomenon is known as fluorescence anisotropy and has generated
a broad literature (see \cite{lakowicz} and \cite{valeur} for reviews).
This effect was used for instance in Organic Light Emitting Diodes, aligning the molecules to optimize the fluorescence emission \cite{2009 Yokoyama,JMC-alignement-OLED}, or in thin films to engineer the second harmonic generation and the two-photon fluorescence \cite{bidault}.
Here, we focus on its consequences on stimulated emission and gain.
Theories were developed to account for amplification anisotropy \cite{yaroshenko,reyzer-1,reyzer-2,Liang Rh},
however we will show hereafter, that it can be described in the framework
of fluorescence anisotropy, in good agreement with experiments.

We experimentally investigated the influence of the pump polarization
on ASE intensity and lasing thresholds with the typical samples and the
characterization setup, which are described in
Section \ref{sec:experience-usual}. ASE experiments were carried out on doped PMMA layers, without lithography step. As in the previous Section, we consider testbed Fabry-Perot microlasers. The direction of observation remains along the width
of the ribbon ($y$-axis in Fig.\ref{fig:setup}b), while the linear
pump beam polarization lies in the substrate plane and is varied between
perpendicular ($\alpha=90^{\circ}$) and parallel ($\alpha=0^{\circ}$)
to the $y$-axis. Fig.\ref{fig:ase-FP} presents the emitted intensity
versus the pump intensity in ASE and Fabry-Perot configurations. It
suggests, that the pump polarization $\alpha$ is a relevant parameter,
which influence is strongly related to the molecular structure of
the dye. For instance, the laser threshold is reduced by a factor
of three from $\alpha=0^{\circ}$ to $90^{\circ}$ for DCM (Fig.\ref{fig:ase-FP}b),
which can be considered as a linear dipole due to its elongated molecular
structure, while the ASE curves remain almost unmodified for PM605
(Fig.\ref{fig:ase-FP}a), which features a more rounded-off skeleton
(see Fig.\ref{fig:spectra}).\\

\begin{figure}[htb]
\centerline{\includegraphics[width=8.3cm]{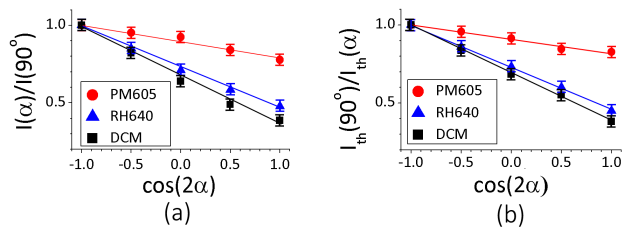}}
\caption{Role of the pump polarization on stimulated emission. (a) ASE in dye-doped
PMMA thin film. (b) 180 $\mu m$ Fabry-Perot micro-lasers. The pump
beam polarization $\alpha$ is defined on Fig.\ref{fig:setup}.b.}
\label{fig:resultats}
\end{figure}

To predict the dependance on $\alpha$, we use a model based on fluorescence
anisotropy \cite{valeur,brasselet,SPIEnous}, which accounts for a
given distribution $f$ of fixed and non-interacting dyes. The emitted
intensity $I_{e}$ is proportional to an integral over all the possible
orientations $\Omega$ of a molecule:
\begin{equation}
I_{e}(\alpha)\propto \int_{\Omega}P_{abs}(\alpha,\Omega)P_{em,y}(\Omega)f(\Omega)d\Omega\label{eq:integrale}
\end{equation}
Here $P_{abs}$ stands for the probability of pump absorption and
is then proportional to a cosinus squared between the pump polarization
and the absorption transition dipole of the molecule. Likewise
$P_{em,y}$ stands for the emission probability in the $y$ direction
and is then a sinus squared between the $y$ direction and the transition
emission dipole of the molecule. Hence we must introduce the angle
$\beta$ between the absorption and emission transition dipoles of
the molecule, which is known to be a constant, depending only on the
molecular structure of the dye \cite{Selenyi}.\\

To calculate expression (\ref{eq:integrale}), we must choose an appropriate
distribution $f$. For instance, $f$ is constant for an isotropic
distribution of molecules. As evidenced by ellipsometry measurements,
spin-coating slightly aligns the molecules in the layer plane. So
we introduce an angle $\theta_{0}$, such that $f$ is constant between
$-\theta_{0}$ and $+\theta_{0}$ and zero outside (see Fig.\ref{fig:theta0}).
A more comprehensive study is reported elsewhere \cite{SPIEnous}.
Here we skip the details and go directly to the following formula \cite{note:erreur-spie-pola}:
\begin{equation}
\frac{I_{e}(\alpha)}{I_{e}(\alpha=90^{o})} =\frac{\rho+1}{2}+\frac{\rho-1}{2}\cos2\alpha,\label{eq:I-de-alpha}
\end{equation}
where
\begin{equation}
\rho=\frac{I_{e}(0^{o})}{I_{e}(90^{\circ})}
\end{equation}
is a positive function of $\theta_{0}$ and $\cos^{2}\beta$, and
is described in Appendix \ref{sec:rho}. As expected, the emitted
intensity $I_{e}$ is maximal for $\alpha=90^{\circ}$, which was
the configuration considered for VSL and threshold measurements (Sec.\ref{sec:vsl}
and \ref{sec:seuil}). For a linear dipole ($\beta=0$) and an isotropic
3D distribution ($\theta_{0}=90^{\circ}$), then $\rho=1/2$, which
means, that a significant part of the light is emitted in the $y$-direction,
even if the pump polarization is parallel to the $y$-direction.\\

$I_{e}(\alpha)$ describes the probability of fluorescence emission.
Hence it is likely that the ASE intensity is proportional to $I_{e}$.
First we checked that the ratio $\frac{I_{e}(\alpha)}{I_{e}(90^{o})}$
does not significantly depend on the pump intensity, namely the fluctuations
remain in the range of $\pm1\%$ for pump intensity varying from 4
to 80 MW.cm\textsuperscript{-2}. The average ratio $\frac{I_{e}(\alpha)}{I_{e}(90^{o})}$
is then plotted versus $\cos(2\alpha)$ in Fig.\ref{fig:resultats}a
for the three dyes in ASE regime. The curves present a linear behavior,
evidencing the validity of formula (\ref{eq:I-de-alpha}), even in
the stimulated emission regime. The $\rho$ values presented in Table
\ref{tab:rho} are then inferred from the linear fits and formula
(\ref{eq:I-de-alpha}).

\begin{table}
{\small{{\centering \caption{{\small{{\label{tab:rho} Comparison of $\rho$ values obtained from
ASE and laser threshold}}}}
}}{\small \par}
{\small{}}}{\small \par}
\centering{}%
\begin{tabular}{cccc}
 &  &  & \tabularnewline
\hline
{\small{{
}}}$\rho$  & {\small{{DCM}}}  & {\small{{PM605}}}  & RH640\tabularnewline
\hline
{\small{{ASE}}}  & 0.33$\pm$0.03  & 0.77$\pm$0.03  & 0.46$\pm$0.03\tabularnewline
Threshold  & 0.38$\pm$0.03  & 0.84$\pm$0.03  & 0.48$\pm$0.04\tabularnewline
\hline
\end{tabular}
\end{table}

Validity of Exp. (\ref{eq:I-de-alpha}) can be checked for lasing
thresholds as well. In fact they can be considered as working points
where the non-linear behavior is still relatively low and fluorescence
formula like Eq.(\ref{eq:integrale}) should then apply \cite{brasselet,casperson}.
The gain value $g_{th}$ necessary to reach the lasing threshold in
the stationary regime is determined by losses, which do not depend
on $\alpha$, leading to $\Gamma K(\alpha)I_{th}(\alpha)=g_{th}=const$.
Thereby, we expect the ratio of threshold intensities $I_{th}(\alpha)/I_{th}(90^{o})$
of the ribbon-shaped micro-lasers to be inversely proportional to
$K(\alpha)/K(90^{o})$. Besides, the emission intensity $I_{e}(\alpha)$
should be proportional to $K(\alpha)$, at least in the linear regime
close to threshold. Therefore $K(\alpha)/K(90^{o})$ should be proportional
to the right part of Eq. (\ref{eq:I-de-alpha}), as well as $I_{th}(90^{o})/I_{th}(\alpha)$.
This latter ratio is plotted versus $\cos(2\alpha)$ in Fig.\ref{fig:resultats}.b
for $w$=180 $\mu$m Fabry-Perot micro-lasers and shows indeed a linear
behavior for the three dyes. So a $\rho$ value can be inferred for
each of the dyes from the right part of Eq. (\ref{eq:I-de-alpha}).
The results are gathered in Tab.\ref{tab:rho} where the error bars
correspond to the fluctuations for ribbon widths varying from 150
to 200 $\mu$m.

Experimental results, presented in this section, were obtained under
linear pump beam polarization. In fact, emission under circularly-polarized
excitation can be described in the same terms based on Eq. (\ref{eq:I-de-alpha}).
In fact, integration over $\alpha$ provides $\frac{I(circular)}{I(\alpha=90^{o})}=\frac{\rho+1}{2}<1$,
implying that a circularly-polarized pump is less efficient than a
linearly-polarized one with $\alpha=90^{o}$. This ratio was verified for
a 165 $\mu$m DCM ribbon micro-laser and gave a $\rho$ value identical
to that in Tab.\ref{tab:rho}.

For the sake of completeness, $\rho$ values should be inferred as well in
the fluorescence regime. However the high doping rate of the layers
favors energy transfer from an excited dye molecule to a neighbour dye
molecule and tends to make the emission isotropic \cite{lefloch}.
Such a tendancy towards isotropy ($\rho=1$) does not occur obviously in our
stimulated systems (ASE and laser). Actually energy transfer is prevented,
since its characteristic time, typical of spontaneous processes, is
much longer than stimulated emission. The different time scales are
evidenced in App.\ref{sec:lifetime}.

To summarize, the $\rho$ parameter quantifies the sensitivity to
polarization. It is specific of a dye molecule and its distribution
in the layer. Such agreement between the $\rho$ values inferred from
both ASE and lasing experiments for the three dyes stresses the validity
of our assumptions as well as the interest of such an approach for
predictions of gain properties. In practice, the pump polarization
must be carefully controlled to optimize the gain of an organic layer,
up to a factor of three.

\section{Spectral features}

\label{sec:Mazumder}

Laser spectra from organic-based devices are in general multimode
and inscribed in an envelope (see Fig.\ref{fig:spectra}). The distribution
of the lasing frequencies is mostly determined by the resonator shape
and is discussed elsewhere \cite{Bogomolny trace 2}. Here, we focus
on the envelope, in particular its central wavelength, which depends
on the gain medium and is relevant for designing an actual device.

\begin{figure}[htb]
\centerline{\includegraphics[width=8.3cm]{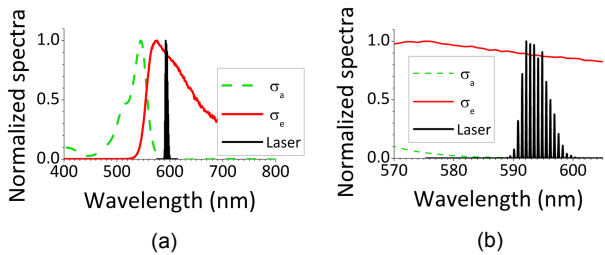}}
\caption{ Absorption and emission properties for PM605. (a) Normalized absorption
cross-section $\sigma_{a}$, normalized stimulated emission cross-section
$\sigma_{e}$, and normalized laser spectrum of a 150 $\mu$m Fabry-Perot
micro-laser. (b) Zoom of (a).}
\label{fig:abs_em_laser}
\end{figure}

Fig.\ref{fig:abs_em_laser} evidences that the envelope of the lasing
spectrum is not centered at the maximum of the fluorescence spectrum.
Here, following Mazumder et al. \cite{Mazumder}, we propose a simple
explanation based on re-absorption. Due to the overlap between absorption
$\sigma_{a}\left(\lambda\right)$ and stimulated emission $\sigma_{e}\left(\lambda\right)$
cross sections, unexcited molecules can absorb photons emitted from
excited states, which decreases the gain:

\begin{equation}
g\left(\lambda\right)=  \sigma_{e}\left(\lambda\right)N^{*}-\sigma_{a}\left(\lambda\right)\left(N-N^{*}\right)
\end{equation}
where \textit{N} stands for the total density of dye molecules and
$N^{*}$ for the density of excited dye molecules. Hence, the ratio
of molecules $\gamma(\lambda)=N_{th}^{*}/N$ that must be excited
to reach the threshold depends on re-absorption \cite{peterson,Mazumder}:

\begin{equation}
\gamma(\lambda)=\frac{N_{th}^{*}}{N}=\frac{g_{th}(\lambda)+\sigma_{a}N}{\sigma_{e}N+\sigma_{a}N}\label{eq:gamma}
\end{equation}

\begin{figure}[htb]
\centerline{\includegraphics[width=8.3cm]{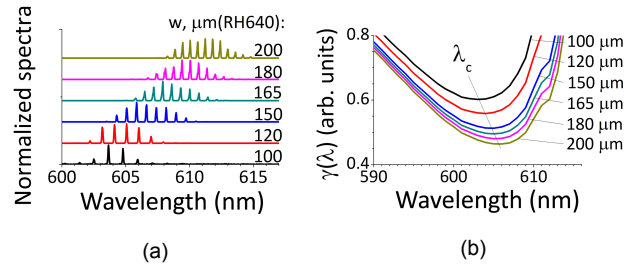}}
\caption{Influence of losses on the spectral envelope: (a) Experimental spectra
of Fabry-Perot cavities of different widths $w$ under the same pump
intensity. The experiments were carried out with a usual sample made
of RH640 and the usual setup, as described in Sec.\ref{sec:experience-usual}.
(b) Threshold condition $\gamma(\lambda)$ calculated with Eq.(\ref{eq:gamma})
for Fabry-Perot cavities of different widths.}
\label{fig:mazumder}
\end{figure}

At threshold, in the stationary regime, the gain balances the losses,
and $g_{th}\left(\lambda\right)$ can be substituted by $-\ln R/\Gamma w$
for a Fabry-Perot resonator (see Eq. \ref{eq:gain-pertes}). $\gamma(\lambda)$
was plotted in Fig.\ref{fig:mazumder}b for different widths $w$
of the Fabry-Perot ($\ln R$ and $\Gamma$ remain unchanged) using $\sigma_{e}$
and $\sigma_{a}$ determined in App.\ref{sec:section-efficace}. Each
curve shows a minimum, which corresponds to the lasing wavelength
close to threshold. The minimum of $\gamma$ is blue-shifted when
the width of the Fabry-Perot decreases, i.e. when the loss of the
cavity increases. The order of magnitude is consistent with the experimental
observations summarized in Fig.\ref{fig:mazumder}a. A similar effect
was reported using absorbers in micro-droplets \cite{Taniguchi,Mazumder}.\\
 In other words, the envelope of the lasing spectrum is blue-shifted,
when the quality factor of the resonator decreases \cite{lam}. Measurements
of such spectral shifts for a given gain material would provide a
solid basis for the experimental estimation of the cavity properties
through the spectroscopic study of the laser effect. Anyway, as such
shift can exceed a dozen of nanometers, it must be taken into account
to optimize the architecture of an actual device.

\section{Measurement of the material gain $g_{mat}$}

\label{sec:pompe-sonde}

The ultimate and most direct way to know the material gain $g_{mat}$ is to
measure the amplification of the layer in a simplified geometry. Therefore in
this Section, we report the measurement of the material gain $g_{mat}$
by the means of a pump-probe experiment, and then its estimation based
on intrinsic characteristics of the gain medium.

\subsection{Pump-probe experiment}

\begin{figure}[htb]
\centerline{\includegraphics[width=7cm]{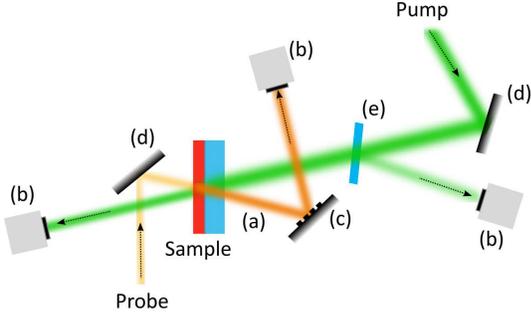}}
\caption{Pump-probe set-up for measuring the material gain $g_{mat}$. Continuous
probe beam and pulsed pump beam. (a) Amplified probe beam. (b) Photodiodes.
(c) Monochromator. (d) Mirrors. (e) Glass slide.}
\label{fig:pompe-sonde-schema}
\end{figure}

\begin{figure}[htb]
\centerline{\includegraphics[width=8.3cm]{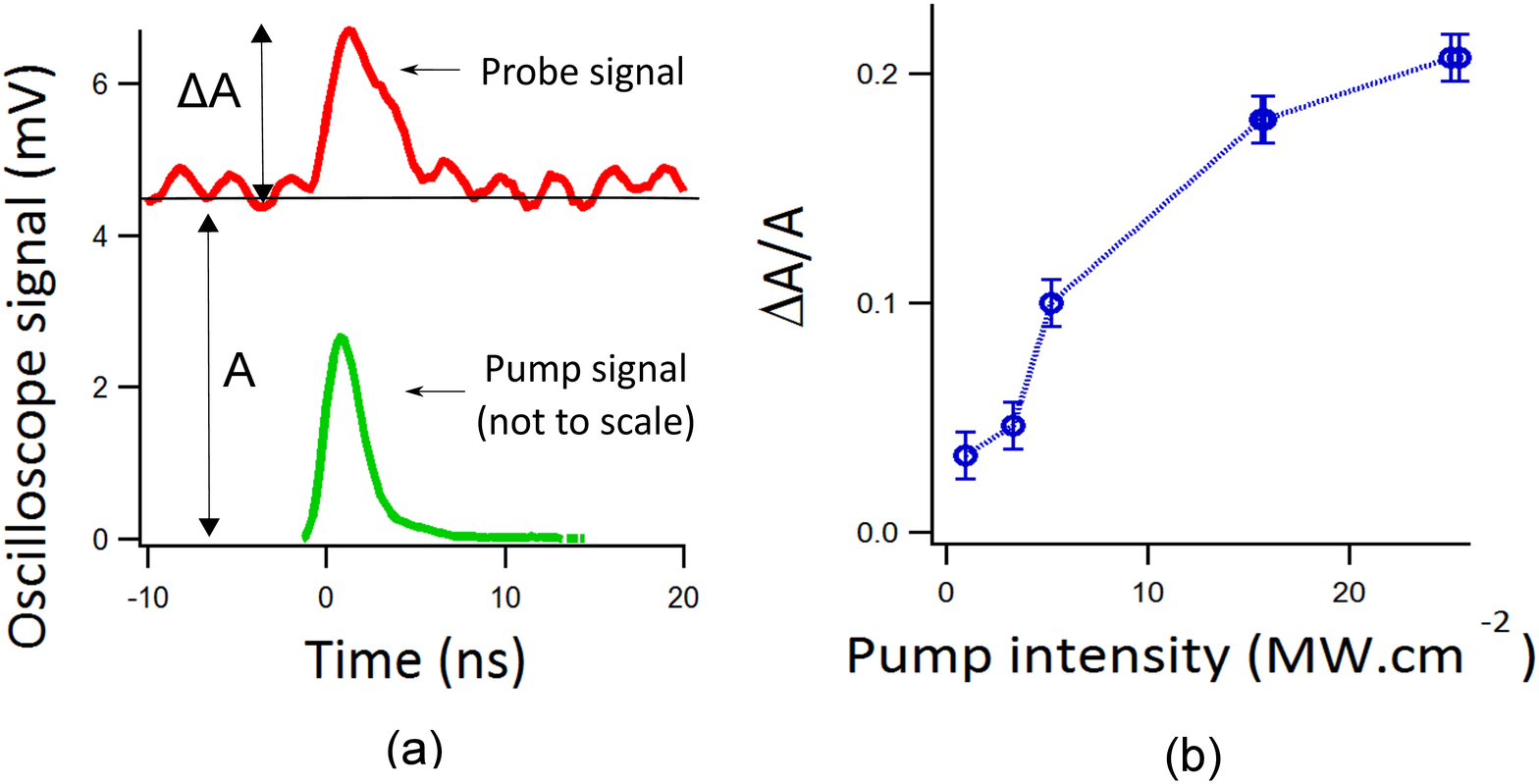}}
\caption{(a) Typical signals of the probe and pump beams detected by the photodiodes.
(b) Amplification factor versus the pump intensity for a probe intensity
$I_{pr}=435$ W.cm$^{-2}$.}
\label{fig:pompe-sonde-bump}
\end{figure}

We carried out pump-probe experiments to get the amplification factor,
and then $g_{mat}$. The set-up is summarized in Fig.\ref{fig:pompe-sonde-schema}.
In this configuration, the pump and the probe beams propagate roughly
perpendicular to the sample plane. As described in Sec.\ref{sec:experience-amplification},
the sample was similar to the usual ones, but optimized for amplification:
(i) the gain layer was 18 $\mu$m thick to increase the amplification
path, and (ii) the concentration of DCM was set to 1.4 wt\% to distribute
the absorption of the pump over the whole thickness. The pump beam
was the same as for microlasers tests (532 nm, 10 Hz, 0.5 ns)
with a diameter of 200 $\mu$m FWHM on the sample, measured with an
imaging system. The probe beam is from a Helium/Neon laser (594 nm,
continuous, $\varnothing$ 65 $\mu$m). Both beams are incident on
the sample at the Brewster angle to avoid parasitic reflections. They
are not exactly parallel for experimental convenience, but their linear
polarizations are parallel. The probe is collected on a rapid photodiode
(rise time about 1 ns), its signal being sent to a 1 GHz oscilloscope
triggered by the pump. A bump is visible, when the pump is on. A typical
trace is presented in Fig.\ref{fig:pompe-sonde-bump}a. Several tests
were performed to ensure that the bump rightly comes from the probe
amplification. In particular, a monochromator showed that there is
no bump out of the probe wavelength (ie. 594 nm). Therefore the bump
does not originate from fluorescence, but from stimulated emission.

To infer the material gain $g_{mat}$ and the gain efficiency $K$
from the amplification factor $\Delta A/A$ measured experimentally
(Fig.\ref{fig:pompe-sonde-bump}), we use the equation of propagation
for the probe intensity $I_{pr}$:
\begin{equation}
dI_{pr}=g_{mat}I_{pr}\, dz=KI_{p}I_{pr}\, dz
\end{equation}
which depends on the pump intensity $I_{p}$. Assuming that the absorption of
the probe is not saturated and remains linear within the thickness
$h$ of the layer - a more realistic situation is treated hereafter - then:
\begin{equation}
\frac{I_{pr}(h)}{I_{pr}(0)}=\frac{A+\Delta A}{A}=e^{KI_{p}h}
\end{equation}
As the amplification factor $\Delta A/A$ is smaller than 1 (see
Fig.\ref{fig:pompe-sonde-bump}a), then, in a linear approximation,
it is proportional to the gain efficiency $K$:
\begin{equation}
\frac{\Delta A}{A}=KI_{p}h\label{eq:amplification}
\end{equation}
In practice, the absorption of the pump beam must be taken into account.
Similar considerations lead to the conclusion that the thickness $h$ of
Eq. (\ref{eq:amplification}) must be replaced by an effective one:
$$
h\rightarrow h_{eff}=\frac{1-e^{-\alpha h}}{\alpha}
$$
where $\alpha$ is the absorption coefficient of the material at the
pump wavelength: $\alpha=N\sigma_{a}(\lambda=532nm)$. With the parameters
of this experiment, the correction corresponds to the absorption of
the pump over an effective thickness $h_{eff}=9$ $\mu$m, ie. half
the actual layer thickness.

\begin{figure}[htb]
\centerline{\includegraphics[width=1\linewidth]{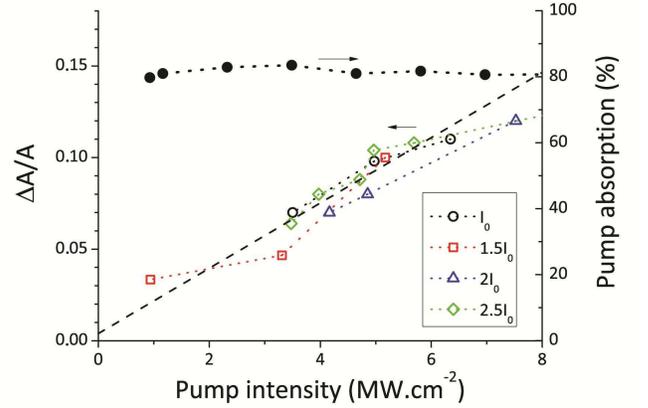}}
\caption{Amplification factor $\Delta A/A$ (left axis) and absorption (right
axis) versus pump intensity. The dotted line is a linear fit of the
set of amplification factors measured at different probe intensities, with $I_0=291$ W.cm$^{-2}$.}
\label{fig:pompe-sonde-resultat}
\end{figure}

As the temporal behavior of the amplification reproduces the temporal
profile of the pump (see Fig.\ref{fig:pompe-sonde-bump})a, the variation
$\Delta A$ was measured at the maximum of the bump signal,
and the factor $I_{p}$ involved in Eq. (\ref{eq:amplification})
is the peak intensity of the pump at the entrance of the layer. The
measured amplification factor $\Delta A/A$ versus the peak intensity
$I_{p}$ is plotted in Fig.\ref{fig:pompe-sonde-resultat} for different
probe intensities. The whole set of experimental data is linearly
fitted, the slope being proportional to the gain efficiency $K$,
leading to $K=20\pm5$ cm.MW$^{-1}$. Here the main source of uncertainty
comes from the temporal profile of the pump beam (assumed to be gaussian)
and its duration.

In summary, amplification measurements with a 1.4 wt\% DCM doped layer
give $K\simeq20$ cm.MW$^{-1}$, while measurements by VSL technique
(Sec.\ref{sec:vsl}) and laser thresholds (Sec.\ref{sec:seuil}) with
a 5 wt\% DCM doped layer lead to $K\simeq80$ cm.MW$^{-1}$. A linear extrapolation from
1.4 wt\% to 5 wt\% gives a slightly different value, $K\simeq70$ cm.MW$^{-1}$, which is consistent with the precision of the measurements.

\subsection{Estimation of the gain efficiency}

In addition, the gain efficiency $K$ can be predicted from the material
characteristics, which is \emph{a priori} more convenient than carrying
out experiments. However, we show in this Section that predictions
significantly overestimate the gain.

The material gain at the wavelength 594 nm is given by
$$
g_{mat}^{594}=\sigma_{e}^{594}N^{*}
$$
where $N^{*}$ is still the density of excited molecules. The vibrational
states are known to relax in less than 1 ps \cite{lakowicz,valeur},
therefore much faster than the timescale involved in this study. We assume
then that the dye molecules can be considered as an effective two
level system, where $N_{0}$ is the density of dye molecules in the
ground state and $N=N_{0}+N^{*}$ is the density of all the dye molecules.
The triplet states are neglected, since the timescale involved in
this study (see App.\ref{sec:lifetime}) is much shorter than the
typical time for intersystem crossing. The rate equation for $N^{*}$
follows then:
\begin{equation}
\frac{dN^{*}}{dt}=\sigma_{a}^{532}N_{0}\frac{I_{p}}{h\nu_{p}}+\sigma_{a}^{594}N_{0}\frac{I_{pr}}{h\nu_{pr}}-\sigma_{e}^{594}N^{*}\frac{I_{pr}}{h\nu_{pr}}-\frac{N^{*}}{\tau_{f}}\label{eq:taux}
\end{equation}
where $\tau_{f}$ is the lifetime of fluorescence. Laser threshold
experiments reported in Sec.\ref{sec:seuil} are consistent with the
assumption of a stationary regime. Hence we still consider this assumption
here, and the left part of Eq. (\ref{eq:taux}) is cancelled out. Then $N^{*}$
can be expressed as:
\begin{equation}
N^{*}=N\frac{\sigma_{a}^{532}\frac{I_{p}}{h\nu_{p}}+\sigma_{a}^{594}\frac{I_{pr}}{h\nu_{pr}}}{\sigma_{a}^{532}\frac{I_{p}}{h\nu_{p}}+\sigma_{e}^{594}\frac{I_{pr}}{h\nu_{pr}}+\frac{1}{\tau_{f}}}
\end{equation}
The cross sections are evaluated in App.\ref{sec:section-efficace}.
The second term in the numerator of $N^{*}$ is several order of magnitude
smaller than the first one, and is thus neglected. Similarly, as $I_{pr}/I_{p}\sim10^{-4}$,
the second term in the denominator is neglected, leading to:
\begin{equation}
N^{*}=N\frac{\sigma_{a}^{532}\frac{I_{p}}{h\nu_{p}}}{\sigma_{a}^{532}\frac{I_{p}}{h\nu_{p}}+\frac{1}{\tau_{f}}}\label{eq:N-excitees}
\end{equation}
The linear regime presented in Fig.\ref{fig:pompe-sonde-resultat}
corresponds to the case:
\begin{equation}
\sigma_{a}^{532}\frac{I_{p}}{h\nu_{p}}\ll\frac{1}{\tau_{f}}\label{eq:saturation-limite}
\end{equation}
which means that $I_{p}$ should be less than 8 MW.cm$^{-2}$, in agreement
with Fig.\ref{fig:pompe-sonde-bump}b. In the linear regime, expression
(\ref{eq:N-excitees}) leads to the following expression for the gain
efficiency:
\begin{equation}
K=\sigma_{e}^{594}\frac{N^{*}}{I_{p}}=\sigma_{e}^{594}\sigma_{a}^{532}N\frac{\tau_{f}}{h\nu_{p}}\label{eq:k-lineaire}
\end{equation}
In App.\ref{sec:aggregats}, $\tau_{f}$ was measured and is about
2 ns. The other quantities involved in expression (\ref{eq:k-lineaire})
are estimated in App.\ref{sec:section-efficace}, and lead to $K\simeq2\,000$
cm.MW$^{-1}$, so 100 times higher than the experimental value. \\
 This estimation leads to two major conclusions. (i) The derivation
of Expression (\ref{eq:k-lineaire}) was not specific to the DCM dye.
Moreover, the numerical values used to estimate $K$ correspond to
the properties of DCM, but do not change significantly from one dye
to another. Hence, the estimation of the gain efficiency $K\sim2\,000$
cm.MW$^{-1}$ is roughly independent of the laser dye, under
the assumption of a stationary and linear regime. (ii) This strong
discrepancy between the estimated and measured gain efficiencies may
be explained by different factors. In fact, several loss processes
were not taken into account in the derivation of Expression (\ref{eq:k-lineaire}),
like for instance absorption by excited states, re-absorption, and
quenching, whereas they were evidenced in Sec.\ref{sec:Mazumder}
and App.\ref{sec:aggregats}.

Furthermore, numerical simulations based on the Tang-Statz-deMars
rate equations \cite{baev,seb-simul} lead to a gain which depends significantly
on the pump duration and the involved losses. In general, the
assumption of stationary pumping overestimates the gain efficiency.
However the actual dynamics and loss processes are difficult to quantify
and strongly depend on the specific thin film which is investigated.
For instance, the presence of aggregates - which leads to quenching
- is strongly dependent on the fabrication process \cite{macromolecule-agregats}.\\

We conclude this Section by pointing out that the material gain $g_{mat}$ can be effectively measured by the means of a pump-probe setup, leading
to a gain efficiency of $K=20$~cm.MW$^{-1}$ at 594 nm for 1.4 wt\%
of DCM in PMMA pumped at 532 nm. This result is in good agreement with the
experimental values inferred from the VSL method in Sec.\ref{sec:vsl}
and laser thresholds in Sec.\ref{sec:seuil}. The predicted $K$ is
two orders of magnitude higher. However this estimation must be treated
with care, since it does not include the dynamics and intra- and inter-molecular
processes which reduce the gain significantly and cannot be easily
quantified.

\section{Conclusion}

In this paper, we have investigated the amplification properties of
dye-doped polymer thin films. The gain efficiency $K$ was introduced
to facilitate the comparison between different materials in various
configurations. It represents the linear ratio between gain and pump
intensity, in the limit of validity of such a linear behavior. It
was consistently measured by different methods (pump-probe amplification
experiment, VSL technique, laser thresholds), and is about
80 cm.MW$^{-1}$ around 600 nm for a 5\% DCM-doped PMMA layer pumped at
532 nm with 0.5 ns pump pulses. A rough
theoretical prediction overestimates its value by two orders of magnitude.
Refinements of this model lead to gain values in better agreement,
but they depend on different features of the dye-doped thin film which
can hardly be measured quantitatively.\\
 Several parameters alter the bulk gain and must be taken into account
to design an actual device. First, the confinement factor $\Gamma$
includes the geometrical features and the overlap between the pump
beam and the propagating laser field. It varies typically from 0.3
to 0.8. Then, the intrinsic anisotropy of the molecular structure induces
a sensitivity to the polarization of the pump beam, and can modify
the gain up to a factor of three. Incidentally, we showed that the gain
measurement is sensitive to the anisotropic distribution of the dye
molecules within the layer, and provides therefore an indirect method
of estimation. Finally, the gain spectrum is monitored by reabsorption
processes and its central wavelength can be predicted relatively well.
For all these parameters, we performed relevant experiments which are mutually consistent and also in agreement with theoretical predictions.\\
 In conclusion, we showed that a theoretical prediction of gain is
in general not reliable. In fact, the actual gain depends strongly
on quenching and on the anisotropy of the dye distribution, which
are monitored by the fabrication process of the layer. Therefore,
we propose to evaluate the gain properties by comparison with well-known
calibrated devices, such as the Fabry-Perot microlasers
reported here. We hope that this work will help to pave the way towards
an electrically pumped hybrid organic-inorganic laser.

\section*{Acknowledgments}

The authors acknowledge J. Delaire, F. Bretenaker, C. Lafargue, and
J. Lautru for stimulating and fruitful discussions. We feel particularly grateful to Kenneth D. Singer for pointing
reference \cite{macromolecule-agregats} and to Alexander Nosich for
his suggestion of Eq. (\ref{eq:confinement-general}).

\appendix

\section{Temporal behavior}

\label{sec:lifetime}

\begin{figure}[htb]
\centerline{\includegraphics[width=7cm]{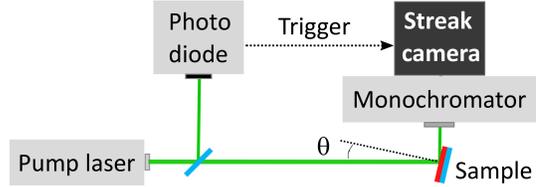}}
\caption{Scheme of the setup used for temporal studies. The retardation line
is not shown. For ASE and Fabry-Perot measurements, $\theta$ is set
to zero, and the emission is collected in the plane of the layer.
For fluorescence measurements, $\theta=55^{\circ}$ in order to prevent
propagation effects.}
\label{fig:streak-schema}
\end{figure}

This Appendix focuses on experimental studies of the temporal properties
of spontaneous and stimulated emission in dye-doped polymer thin films.
The usual dye-doped PMMA layers described in Sec. \ref{sec:experience-usual} were
pumped with a frequency doubled Nd:YAG laser (10 Hz, 532 nm, 35 ps)
and the emission was collected through a monochromator and then injected
into a streak camera (Optoscope by ARP) with about 8 ps temporal resolution.
Fig.\ref{fig:streak-schema} presents a simplified scheme of the setup.
In order to prevent the influence of guiding effects, the layers used
for fluorescence study were directly spin-coated on a Si substrate
and $\theta$ was set to $55^{\circ}$, while a SiO$_{2}$/Si substrate
was used for ASE and Fabry-Perot samples with $\theta=0^{\circ}$.\\

\begin{figure}[htb]
\begin{minipage}[c]{0.48\linewidth}
\includegraphics[width=1\columnwidth]{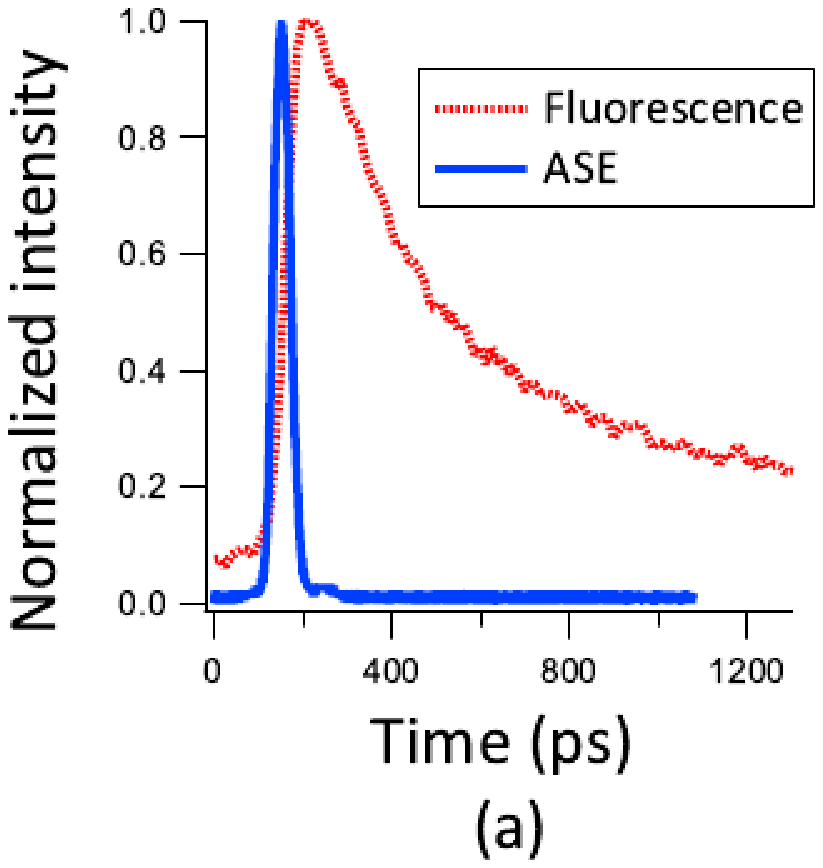}
\end{minipage}
\begin{minipage}[c]{0.48\linewidth}
\includegraphics[width=1\columnwidth]{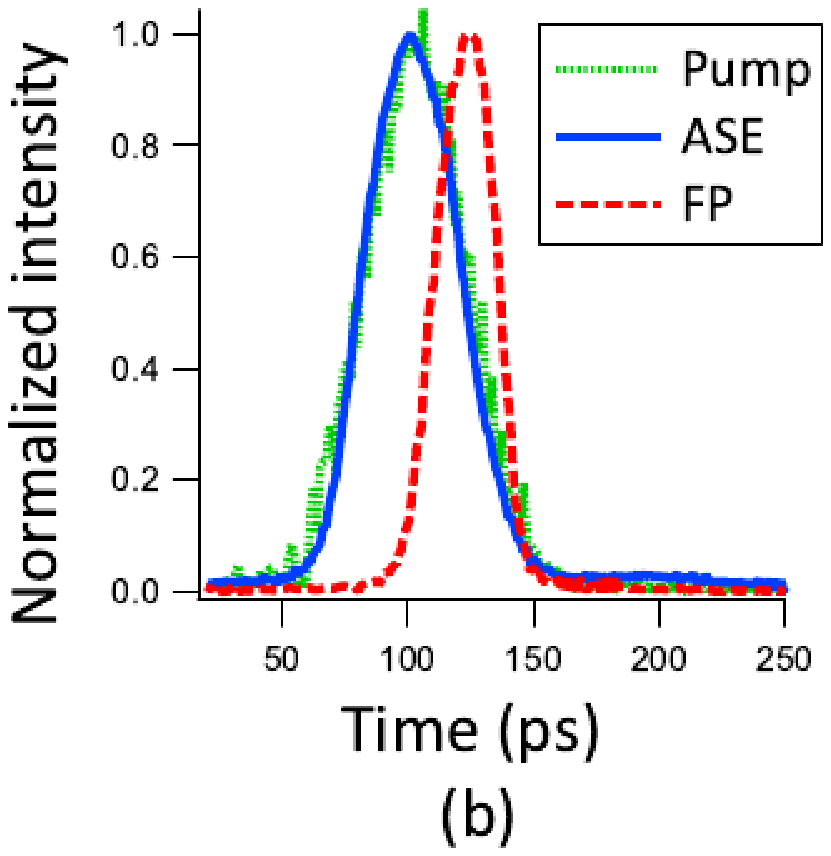}
\end{minipage}\caption{Dynamics of the spontaneous and stimulated emissions in a PMMA layer doped with 5 wt\% of DCM: (a) fluorescence and ASE ,
(b) Pump signal, ASE, and lasing emission from a Fabry-Perot cavity
of width $w=200$ $\mu$m just above the threshold. Averaging over
20-100 pump pulses.}
\label{fig:ase-fluo-lifetime}
\end{figure}

As evidenced in Fig.\ref{fig:ase-fluo-lifetime}, the temporal behavior
of the ASE signal replicates that of the pump (except for RH640, which
exhibit a small exponential relaxation of ASE, not shown here), whereas
the fluorescence emission occurs over a longer time scale.
Regarding the lasing emission from a Fabry-Perot cavity, Fig.\ref{fig:ase-fluo-lifetime}b
evidences that it is delayed by some 20 ps. This delay decreases
if the pump intensity increases and the Fabry-Perot width $w$ decreases,
as expected since the build-up time of the laser emission decreases
as well. It can be noticed that no spiking is observed at a 100 ps scale, contrary to
\cite{alq3-dcm laser}, maybe due to the shorter pump pulse.

\section{Fluorescence anisotropy}

\label{sec:rho}

\begin{figure}[htb]
\centerline{\includegraphics[width=7cm]{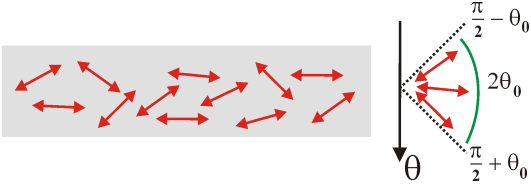}}
\caption{Definition of $\theta_0$. Left: Scheme of a polymer layer with randomly distributed fluorophores (not to scale). Right: $\theta_0$ is defined such as each dye is orientated between $-\theta_0$ and $+\theta_0$.}
\label{fig:theta0}
\end{figure}

This Appendix provides the expression of the $\rho$ factor mentioned
in Sec.\ref{sec:pola}. We consider an ensemble of fixed and non-interacting
fluorophores, and are looking for the emitted intensity in the $y$
direction, $I_{e}$, depending on the polarization of the linear pump
beam $\alpha$ (see Fig.\ref{fig:setup}b for notations). $\rho$
is defined as $I_{e}(\alpha=0)$ normalized by $I_{e}(\alpha=\pi/2)$:
$$
\rho=\frac{I_{e}\left(\alpha=0\right)}{I_{e}\left(\alpha=90\right)}
$$
and was introduced to simplify formula (\ref{eq:I-de-alpha}).\\
 In practice, $\rho$ depends on two parameters, namely the distribution
of dyes and the angle $\beta$ between the absorption and emission
dipole moments, which is roughly constant for a given dye. Here
we chose an isotropic distribution of dyes within an angle $\theta_{0}$,
as depicted in Fig.\ref{fig:theta0}. Then, the limit case $\theta_{0}=\pi/2$
corresponds to a three-dimensional isotropic distribution, and $\theta_{0}=0$
to a bi-dimensional distribution, where the dyes lie within the layer
plane, which most likely applies to spin-coated light-emitting polymers
\cite{polymere orient Ifor}.\\

A comprehensive derivation is described elsewhere \cite{SPIEnous},
the final formula being as follows:
\begin{equation}
\rho=\frac{A_{1}\left(\beta\right)B_{1}\left(\theta_{0}\right)+A_{2}\left(\beta\right)B_{2}\left(\theta_{0}\right)}{A_{1}\left(\beta\right)B_{1}\left(\theta_{0}\right)+A_{2}\left(\beta\right)B_{3}\left(\theta_{0}\right)}\label{eq:annexe-rho}
\end{equation}
where the functions $A_{i}\left(\beta\right)$ and $B_{i}\left(\theta_{0}\right)$
are defined by:
\begin{align*}
A_{1}\left(\beta\right)=  2-2\cos^{2}\beta,\\
A_{2}\left(\beta\right)=  1-3\cos^{2}\beta,
\end{align*}
\begin{align*}
B_{1}\left(\theta_{0}\right)=  \,30+10\sin^{2}\theta_{0},\\
B_{2}\left(\theta_{0}\right)=  -15-10\sin^{2}\theta_{0}+9\sin^{4}\theta_{0},\\
B_{3}\left(\theta_{0}\right)=  -45+10\sin^{2}\theta_{0}+3\sin^{4}\theta_{0}.
\end{align*}
and are introduced for the sake of compacting Eq. (\ref{eq:annexe-rho}).
In limit cases of 2-D ($\theta_{0}=0$) and 3-D ($\theta_{0}=\pi/2$)
distributions, the expression of $\rho$ reduces to:
\begin{align*}
\rho_{2D}\left(\beta\right)=  \frac{3-\cos^{2}\beta}{1+5\cos^{2}\beta}
\end{align*}
\begin{align*}
\rho_{3D}\left(\beta\right)=  2\,\frac{2-\cos^{2}\beta}{3+\cos^{2}\beta}
\end{align*}
For instance, in the case of a 2D distribution and parallel dipole moments
for absorption and emission (ie. $\beta=0$), then $\rho=1/3$. For
DCM, it was reported in \cite{lefloch} that $\beta=25^{\circ}$ under
excitation at 461 nm. Assuming a 2D distribution of chromophores leads
to $\rho=0.43$.

\section{Absorption and stimulated emission cross-sections}

\label{sec:section-efficace}

In this Appendix, the values of the absorption and emission cross-sections
are inferred from experimental data.

\subsection{Absorption cross-section}

The factor $\sigma_{a}(532)N$ is inferred from the absorption of
the pump, see Fig.\ref{fig:pompe-sonde-resultat}:
$$
e^{-\sigma_{a}(532)Nh}\simeq0.2\rightarrow\sigma_{a}(532)N\simeq9.10^{2}\textrm{cm}^{-1}
$$
where $h=18\,\mu$m is the thickness of the layer in the configuration
of Fig.\ref{fig:pompe-sonde-resultat}.

\subsection{Emission cross-section}

The emission cross-section is shaped as the fluorescence spectrum
$E(\lambda)$ and given by the following formula \cite{livre-seb,peterson}:
\begin{equation}
\sigma_{e}(\lambda)=\frac{\phi\lambda^{4}}{8\pi n^{2}c\tau_{rad}}\frac{E(\lambda)}{\int E(\lambda)d\lambda}\label{eq:section-emission}
\end{equation}
where $c$ is the velocity of light in the vacuum. If we assume that
fluorescence is the only decay channel from the $S_{1}$ excited state,
then the quantum yield $\phi$ is equal to 1, and the radiative lifetime
$\tau_{rad}$ is equal to the fluorescent lifetime measured in Sec.\ref{sec:aggregats},
$\tau_{rad}\simeq2$ ns. Formula (\ref{eq:section-emission}) is based
on two main assumptions, which seems valid in this study: (i) the
emission occurs from the lowest vibrational state of the $S_{1}$
band, and (ii) the density of electromagnetic modes is purely classical,
without quantum confinement effect. With a refractive index $n=1.5$,
formula (\ref{eq:section-emission}) leads to:
$$
\sigma_{e}(\lambda=594\,\textrm{nm})\simeq4.10^{-16}\,\textrm{cm}^{2}
$$
It must be emphasized, that the value of $\sigma_{e}$ cannot change
very much from one molecule to another one, since the refractive index
is typically 1.5-2, the radiative lifetime is about 1 ns, and the
fluorescence spectrum is about 50 nm wide. Hence, formula (\ref{eq:section-emission})
leads to more or less the same emission cross-section whatever is
the dye.

\section{Evidences of aggregates and fluorescence lifetimes}

\label{sec:aggregats}

In this Appendix, we report experiments evidencing the presence of
aggregates in the dye-doped layer. Actually these aggregates quench
the emission, and then lead to a decrease of the gain.\\
 Experiments were carried out with different doping rates of DCM in
a PMMA layer spin-coated on a glass slide. The excitation was provided
by a frequency doubled Yb:KGW laser (10 MHz, 400 fs) and the fluorescence
was collected by a time-resolved single-photon counting photo-multiplier
(QA, Europhoton Gmbh). The setup is described in \cite{flim} and
was initially dedicated to the study of the relaxation of fluorescence
anisotropy. Here, we consider only data related to evidences of aggregates.

\begin{figure}[htb]
\centerline{ \includegraphics[width=8cm]{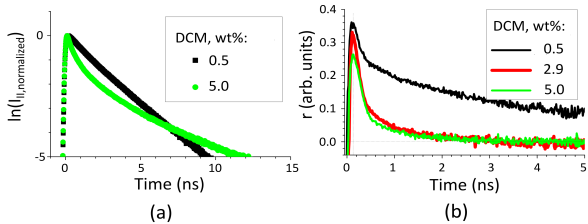}}
\caption{Dynamics of fluorescence for several concentrations of DCM in a PMMA
layer: (a) $I_{||}$ component of the fluorescence intensity. It is polarized
along the polarization of the pump, in contrast to $I_{\bot}$, which is polarized
orthogonally to the pump. (b) Polarization anisotropy parameter $r=\left(I_{||}-I_{\bot}\right)/\left(I_{||}+2I_{\bot}\right)$
a standard fluorescence anisotropy parameter.}
\label{fig:aggregats}
\end{figure}

Figure \ref{fig:aggregats} presents the dynamics of fluorescence
for two dye concentrations. For 0.5 wt\% of DCM, the plot is linear
in logarithmic scale, which corresponds to a monoexponential decay,
as expected for single dye molecules. For 5 wt\% of DCM, the plot
is obviously not linear, which evidences the presence of dye aggregates
in the layer. Moreover the shape of the curve is in good agreement
with the usual theory \cite{pope}. Actually, depending on the geometrical
arrangement of the dye molecules inside an aggregate, the fluorescence rate
is either increased (J-aggregates) or slowed down (H-aggregates).
Indeed, the slope of the curve with 5 wt\% of DCM is first higher
and then lower than the slope of the curve with 0.5 wt\% of DCM, which
evidences the presence of both types of aggregates.\\
This observation is fully consistent with a systematic study of dye
aggregation reported in \cite{macromolecule-agregats}. In fact, to
optimize the optical quality of the dye-doped PMMA layer, it is annealed
at 120$^{\circ}$C, ie. above the T$_{g}$ of PMMA. As reported in
\cite{macromolecule-agregats}, this process ``leads to irreversible
phase separation and the formation of dye aggregates''.\\
In consequence, the fluorescence lifetimes were measured with slightly doped layers (0.5 wt$\%$), and lead to 1.8 ns for DCM, 2.6 ns for RH640 and 1.5
ns for PM605, in consistence with measurements reported elsewhere \cite{bondarev}.


\begin{thebibliography}{10}
\bibitem{note:limite-seuil}
The lowest theoretical limit of the electrically-driven threshold
in organic semiconductor material is about 80 A.cm$^{-2}$ \cite{kozlov cur abs}. It was estimated from the value of the optically-pumped threshold under assumption of maximum recombination rate and neglecting the current-induced losses and metal contact\cite{McGehee}, since the lowest reported threshold under optical pumping is of the order of 100 W.cm$^{-2}$, reached in conjugated polymers \cite{gupta 100W/cm2}.
\bibitem{note:seuil-mesure} There is no significant difference whatever is plotted - maximum of
a spectrum or sum over all the pixels - and how the threshold intensity
is determined from the plot.
\bibitem{note:erreur-spie-pola} Error in Eq. (6) of \cite{SPIEnous}. It should be read $\cos2\varphi\cos2\alpha$ instead of $\cos(2\varphi+2\alpha)$.
\end{thebibliography}
\end{document}